\documentclass[aps,prl,showpacs,twocolumn,a4paper,10pt]{revtex4}
\usepackage{amsmath}
\usepackage{graphicx}
\usepackage{dcolumn}
\usepackage{bm}

\begin{document}

\title{Abnormal Electronic Transport in Disordered Graphene Nanoribbon}
\author{Yan-Yang Zhang$^1$, Jiang-Ping Hu$^2$, X. C. Xie$^{3,1}$, and W. M. Liu$^1$}
\affiliation{$^1$Beijing National Laboratory for Condensed Matter
Physics, Institute of Physics, Chinese Academy of Sciences, Beijing
100080, China.} \affiliation{$^2$Department of Physics, Purdue
University, West Lafayette, Indiana 47907, USA}
\affiliation{$^3$Department of Physics, Oklahoma State University,
Stillwater, Oklahoma 74078, USA}

\date{\today}

\begin{abstract}
We investigate the conductivity $\sigma$ of graphene nanoribbons
with zigzag edges as a function of Fermi energy $E_F$ in the
presence of the impurities with different potential range. The
dependence of $\sigma(E_F)$ displays four different types of
behavior, classified to different regimes of length scales decided
by the impurity potential range and its density. Particularly, low
density of long range impurities results in an extremely low
conductance compared to the ballistic value, a linear dependence
of $\sigma(E_F)$ and a wide dip near the Dirac point, due to the
special properties of long range potential and edge states. These
behaviors agree well with the results from a recent experiment by
Miao \emph{et al.} (to appear in Science).
\end{abstract}

\pacs{72.10.-d, 72.15.Rn, 73.20.At, 73.20.Fz}

\maketitle

\emph{Introduction.}---Recent breakthrough in graphene fabrication
has attracted many attentions to this two-dimensional (2D) material
\cite{7}. The honeycomb lattice structure of graphene gives rise to
two interesting electronic properties in the low energy region which
are distinct from conventional 2D materials, i.e., two valleys
associated to two inequivalent points $K$ and $K'$ at the corner of
the Brillouin zone, and linear ``Dirac-like'' rather than quadratic
bare kinetic energy dispersion spectra. Many of the interesting
experimental results are attributed to these peculiar properties
near Dirac point, the Fermi level for undoped graphene.

Some interesting aspects of the electronic transport in disordered
graphene have been investigated theoretically
\cite{Zie98,Kh06,Al06,At06} and experimentally
\cite{Moro06,Wu07,Hrd07,Ru07}. It was realized that the potential
range of the impurities plays a special role in the electronic
transport in graphene \cite{2,1}. Impurities with long range
potential scattering were considered to be a possible origin for
some unconventional features in the experiments
\cite{3,Mor06,15,16,Hwang07,Wa07}. Such a potential could be
realized by screened charges in the substrate. The peculiarity of
the long-range disorder is the absence of valley mixing due to the
lack of scattering with large momentum transfer. In a realistic
experiment, a gate voltage $V_g$ can continuously tune the carrier
density (thus the Fermi energy $E_F$) in the graphene sample. A
perfect linear relation between conductivity $\sigma$ and gate
voltage $V_g$ was observed \cite{7}. However, clear nonlinear
$\sigma(V_{g})$ curves emerge in a recent experiment
\cite{Miao07}. For large $V_{g}$, the $\sigma(V_{g})$ curves show
a sub-linear behavior, i.e., square root in $V_{g}$ rather than a
linear one. The conductance is smaller than the theoretical
ballistic value by a factor of 3-10. Whereas in the low $V_g$
region near the Dirac point, even this square-root like behavior
breaks down, and a wide dip appears. This dip is wider for a
smaller sample. These observed novel transport features have no
explanations thus far.

\begin{figure}[t]
\includegraphics[bb=0 0 322 226,width=0.4\textwidth]{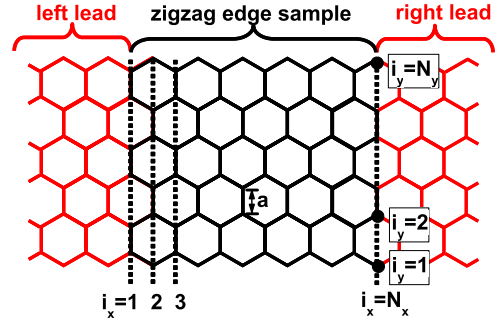}
\caption{(color online) Geometry of the device for measuring the
conductance of graphene nanoribbon. Two semi-infinite leads (red)
are connected to the graphene nanoribbon sample with zigzag edges
(black). The coordinate of each site is labelled as
($i_{x}$,$i_{y}$), where $1 \leq i_{x} \leq N_{x}$ and $1 \leq i_{y}
\leq N_{y}$ are integers.} \label{F1}
\end{figure}

In this Letter, we perform systematic calculations to investigate
the effect of the impurity potential range and its density on the
conductance of graphene nanoribbon. The dependence of
$\sigma(E_F)$ displays four different types of behavior,
corresponding to regimes with different length scales depending on
the range and density of the impurities, which can be used as
criterions in the experiments. Moreover, we demonstrate that the
nonlinear dependence in the recent experiment \cite{Miao07} can
also be explained when scattering due to the low density and long
range impurities are accounted.

\emph{Model and Method.}---We consider a two-terminal device to
calculate the conductance, which includes a graphene nanoribbon and
two leads with zigzag edges, as shown in Fig. \ref{F1}, where the
graphene sample is divided into $N_{x}$ vertical chains with $N_{y}$
sites in each chains. In this setting, the coordinate ($i_{x}$,
$i_{y}$) ($1\leq i_{x}\leq N_{x} $, $1\leq i_{y}\leq N_{y}$) of each
site is labelled. Two clean and semi-infinite leads are assumed to
have the same type of lattice as the graphene sample \cite{6,14} to
avoid additional scattering contribution  from the mismatched
interfaces between different types of lattices.

We describe graphene by the tight binding Hamiltonian for the $\pi$
orbital of carbon \cite{17,Pe06n2}
\begin{equation}
H=\sum_{i}\epsilon_{i}c_{i}^{\dagger }c_{i}+t\sum_{\langle
i,j\rangle }(c_{i}^{\dagger }c_{j}+\text{H.c}),  \label{1}
\end{equation}%
where $c_{i}^{\dag }$ ($c_{i}$) creates (annihilates) an electron
on site $i$, $\varepsilon _{i}$ is the potential energy and $t$
($\sim $2.7eV) is the hopping integral between the nearest
neighbor carbon atoms with distance $a$ ($\sim $1.42{\AA }). We
use $t$ as the energy unit and $a$ as the length unit.

In the presence of disorder, $N_{i}$ impurities are randomly
distributed among $N$ ($\equiv N_{x}\times N_{y}$) sites. The
potential energy of the $i$-th site $\varepsilon _{i}$ at position
$\mathbf{r}_i$ is induced by these impurities as \cite{3,Wa07}
\begin{equation}
\varepsilon _{i}(\mathbf{r}_i)= \sum_{n=1}^{N_{i}}V_{n}
\exp(-|\mathbf{r_i}-\mathbf{r_{n}}|^{2}/(2\xi^{2})), \label{2}
\end{equation}%
where $\mathbf{r_{n}}$ is the position of the $n$-th impurity, $\xi$
represents the spatial range of the impurity potential, and the
potential strength $V_{n}$ of the impurities is randomly distributed
in the range $(-W/2,W/2)$, independently. The average distance
between two impurities $R_i$ can be defined as $R_i\sim \sqrt{L_{x}
L_{y}/N_{i}}$, where $L_{x(y)}$ is the length (width) of the
rectangular sample. We shall show in the following that
 distinct interesting phenomena can be observed when the system is in
the different regimes of these length scales that is determined by
 $L_{x(y)}$, $\xi$ and $R_i$.

\begin{figure}[t]
\includegraphics[bb=0 0 326 241,width=0.4\textwidth]{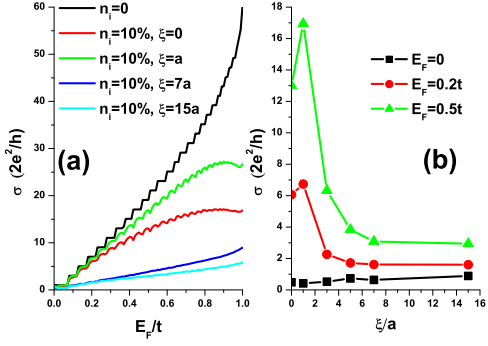}
\caption{(color online) (a) Conductivity $\sigma$ of graphene
nanoribbons with zigzag edges as a function of Fermi energy $E_F$
with different potential range $\xi$. (b) Conductivity $\sigma$ as a
function of potential rnge $\xi$ at different Fermi energy $E_F$.
$N_{x}=106$, $N_{y}=60$ ($L_{x}=90.9a$ and $L_{y}=89.0a$), $W=0.5t$
for both cases, where $t=2.7$eV and $a=1.42$nm. The conductivity is
averaged over 100 random configurations for each curve.} \label{Fxi}
\end{figure}

In the framework of the non-equilibrium Green's function method ,
the zero temperature conductance $G$ and density
of states (DOS) $\rho$ of the sample at Fermi energy $%
E_{F}$ can be written as $G(E_{F})= \frac{2e^{2}}{h}
\text{Tr}(\text{Im}\Sigma _{L}^r(E_{F}) G^{r}(E_{F})\text{Im}\Sigma
_{R}^r(E_{F})G^{a}(E_{F}))$ and $\rho(E_{F}) = -\frac{1}{\pi
}\text{Im}[\text{Tr}G^{r}(E_{F})]$, where $G^{r(a)}(E_{F})$ is the
retarded (advanced) Green's function, and $\Sigma _{L(R)}^{r(a)}$ is
the retarded (advanced) self-energy due to the left (right) lead
 \cite{4,5,6,BKN01}. The conductivity is related to the conductance by the
geometric relation $\sigma =L_{x}G/L_{y}$.

\begin{figure}[t]
\includegraphics[bb=0 0 352 263,width=0.4\textwidth]{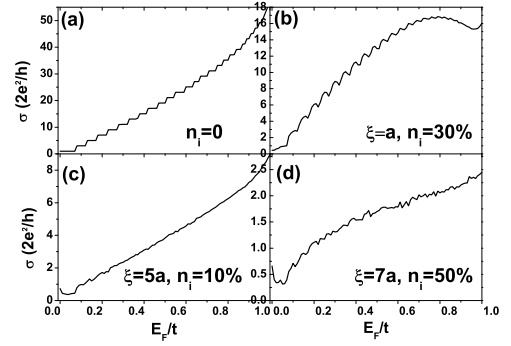}
\caption{Conductivity of graphene nanoribbons with zigzag edges as a
function of Fermi energy, with $N_{x}=106$, $N_{y}=60$
($L_{x}=90.9a$ and $L_{y}=89a$), $W=0.5t$ at different $\xi$ and
$n_{i}$. The conductivity is averaged over 100 random configurations
for each curve. Note different $\sigma$ scalings in each regime. (a)
Regime 1, no impurity. (b) Regime 2, short range impurities. (c)
Regime 3, long range and low density impurities. (d) Regime 4, long
range and high density impurities.} \label{FxiD}
\end{figure}

In the clean limit, a self-consistent calculation for graphene in
the Hartree approximation shows that, $\sigma\propto\sqrt{V_g}$
\cite{Ro07}. In this case, $\sigma \propto E_F$ \cite{18} (also see
Fig. \ref{FxiD} (a)). This leads to $E_F\simeq
\alpha_{V_g}\sqrt{V_g}$, where $\alpha_{V_g}$ is a device-dependent
prefactor, whose typical value $\sim 10^{-3}-10^{-2}$. This relation
between $E_f$ and $V_g$ is valid even in the presence of disorder
since it is a global response. Therefore, we can concentrate on the
relation between $\sigma$ and $E_f$.

\emph{Results and Discussions.}--- Firstly, we investigate the
effect of the potential range $\xi$ and density $n_{i}\equiv N_i/N$
of the impurities with fixed $W=0.5t$. Let us start by having a
first glance at the effect of $\xi$. In Fig. \ref{Fxi} (a), we plot
$\sigma(E_{F})$ with different $\xi$. A direct conclusion from this
figure is that, in most regions of $E_F$, short range ($\xi=0,1a$)
impurities can lead to a considerable decrease of conductivity, but
long range ($\xi=7a,15a$) impurities will decrease the conductivity
much further, in the case of same impurity density. In the
experiments by Miao \emph{et al.} the conductance is smaller than
its ballistic value by a factor of 3--10, in the range of high $V_g$
\cite{Miao07}. This suggests that the samples used in these
experiments must include long range impurities. Near the Dirac point
$E_{F}=0$, this rapid variation breaks down, see Fig. \ref{Fxi} (b).
The magnitude of $\sigma(E_{F}=0)$ is relatively universal ($\sim
e^2/h$), compared to $\sigma$ at finite $E_F$. This is consistent
with the experiment \cite{7}. The linear dependence of $\sigma(E_F)$
obtained in the mean field theory \cite{Zie06} is not entirely
valid.  Our numerical results indicate that $\sigma(E_F)$ is a more
complicated function that depends on the nature of disorder. To
illustrate the physics, we focus on  the conductivity $\sigma$ for
graphene nanoribbons with zigzag edges for different $\xi$ and
$n_{i}$, which can be seen in Fig. \ref{FxiD}. All these behaviors
can be classified into four typical regimes.

\begin{figure}[t]
\includegraphics[bb=0 0 356 265,width=0.45\textwidth]{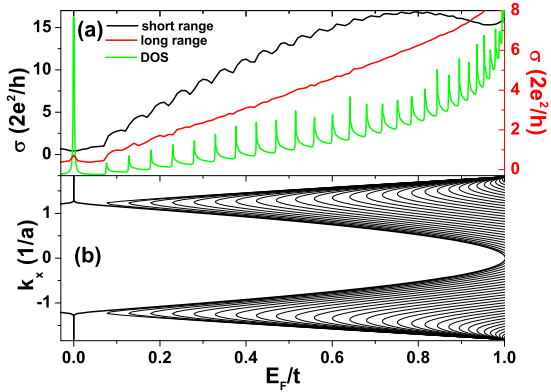}
\caption{(color online) (a) Averaged conductivity of graphene
ribbons with short range impurities (black) (Data are same with
Fig.\ref{FxiD} (b)) and long range impurities (red) (Data are same
with Fig.\ref{FxiD} (c)), and density of states of a clean ribbon
with the same size (green, in arbitrary unit). (b) Band structure of
a zigzag edge graphene nanoribbon with the same size. Note
correspondences between these figures.} \label{FDD}
\end{figure}

\emph{Regime 1, no impurity, $n_{i}=0$.} When there is no impurities
($n_{i}=0$), $\sigma (E_F)$ increases almost linearly except for
small quantized plateaus due to finite size quantization in the
transverse direction \cite{18}, as shown in Fig. \ref{FxiD} (a).
When the sample is large enough, these sub-structures due to finite
size effect can be ignored, so $\sigma(E_F)$ is linear, and $\sigma
\sim \sqrt{V_g}$ according to $E_F\simeq \alpha_{V_g}\sqrt{V_g}$
\cite{Ro07}.

\emph{Regime 2, short range impurities, $\xi \lesssim a$.} When
short range impurities are present (Fig. \ref{FxiD} (b)), the first
distinct feature is the sub-linear behavior of $\sigma (E_{F})$,
especially in the high energy region. This anomaly can be understood
as enhanced scattering due to large level broadening when DOS is
large. In the coherent phase approximation without vertex
corrections, the conductivity $\sigma$ of disordered system can be
written as \cite{20}
\begin{equation}
\sigma=\frac{S_{F}}{8 \pi^{2}}\frac{e^{2}v_{F}}{|\Sigma_{2}|},
\label{5}
\end{equation}%
where $S_{F}$ is the area of the Fermi sphere and $\Sigma_{2}$ is
the imaginary part of self energy corresponding to the disorder
scattering. The level broadening $|\Sigma_{2}|=\pi \rho N_{i}
|V(q)|^{2}$, where $V(q)$ is the Fourier transform of impurity
potential \cite{20}. For a disordered nanoribbon, the van Hove
singularity at $E=t$ of 2D graphene \cite{Naka96,4} degenerates into
a finite but still very sharp peak (see the green curve in
Fig.\ref{FDD} (a)). Therefore, $\Sigma_{2}$ also has a sharp peak at
this point, giving rise to a minimum of $\sigma$ as can be seen from
(\ref{5}). This minimum $\sigma$ at $E_{F}=t$ leads to a sub-linear
$\sigma(E_{F})$ and \emph{sub-square root} $\sigma(V_g)$, according
to the relation between $E_F$ and $V_G$ mentioned above.

Such level-broadening enhanced scattering also happens at the
bottoms of sub-bands, where van Hove singularities emerge
\cite{13,QTD}. Indeed, when disorder is not strong enough to smear
these singularities out completely, a small dip can be observed at
each sub-band bottom, as can be seen from Fig. \ref{FDD} (a).

Variation of $n_i$ of short range impurities does not change the
qualitative behavior of $\sigma(E_F)$, but reduces the magnitude of
$\sigma$ for a given $E_F$, when $\xi \sim1$, as one expects.
However, for extremely short range impurities ($\xi\sim 0$), we find
that even the magnitude of $\sigma(E_F)$ is quite independent of
$n_i$ when $n_i>20\%$.

\emph{Regime 3. long range and low density impurities, $\xi\gg a$
and $\xi\lesssim R_i$.} When the potential range $\xi$ increases
further, interesting physics appears. As can be seen in Fig.
\ref{FxiD} (c), the $\sigma(E_{F})$ curves resume their linear
behavior in most energy regions (while the slope is much smaller
as mentioned above). This manifests suppression of large momentum
scattering due to the long range impurities.

\begin{figure} [t]
\includegraphics[bb=0 0 342 240,width=0.4\textwidth]{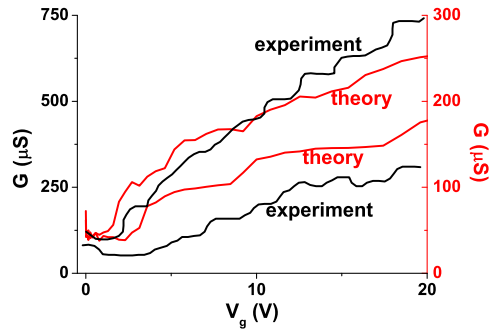}
\caption{(color online) Conductivity $\sigma$ as a function of
$V_g$: A qualitative comparison between theory and experiment. The
black lines represent experimental results by Miao \emph{et al.}
\cite{Miao07}. The upper curve is for a larger sample and the lower
curve is for a smaller sample. The red lines represent our numerical
results. The upper curve: $N_x=120$, $N_y=110$, $\xi=4a$, $W=2t$;
The lower curve: $N_x=72$, $N_y=80$, $\xi=4a$, $W=3t$. The
conductivity is averaged over 50 random configurations for each
curve.} \label{Fcompare}
\end{figure}

Another notable nonlinear $\sigma(E_{F})$ can be observed near the
Dirac point, where a wide dip appears. We find this happens within
the energy region where the first sub-band is visible for a clean
graphene (see Fig.\ref{FDD} (b)). For a smaller graphene sample,
different sub-bands are more separated in energy than for a larger
sample, giving rise to wider quantized conductance plateaus, and
also a wider dip near the Dirac point, which is consistent with the
experimental result \cite{Miao07}.

This wide dip can be understood as follows. As shown in Fig.
\ref{FDD} (b), the band structure near the Dirac point is composed
of two branches of subbands, i.e., the upper band $+ E(k_x)>0$ and
the lower band $-E(k_x)<0$. These branches correspond to binding and
antibinding states localized at different edges and sublattices
\cite{Naka96,Ko06}. The wave functions of these edge states possess
different signs according to two edges, sublattices and branches
\cite{Hi03}. Long range impurities will (while short range ones will
not) couple these edge states, giving rise to rather large
scattering matrix elements between two valleys. When $E=0$, two
branches and valleys degenerate, the magnitude of scattering matrix
elements decreases since different signs of these edge states at
$E=0$ making a larger possibility of canceling each other. This is
verified by numerical calculations for scattering matrix elements.

Therefore, $\sigma(V_g)$ behaves in a square root way in this
regime, by noting $E_F \simeq \alpha_{V_g} \sqrt{V_g}$. In Fig.
\ref{Fcompare}, we plot $\sigma(V_g)$, setting
$\alpha_{V_g}=10^{-3}$. A perfect quantitative fitting cannot be
reached because the size of the sample in the experiments is the
order of $N_{x,y}\sim 10^{3}$, which is well out of the capability
of numerical calculations. But the qualitative features, i.e.,
sub-linearity and wide dip in the experiment (Fig. \ref{Fcompare})
can be clearly seen. Once again, we attribute this experimental
result to the contribution of the low-density and long-range
impurities.

\emph{Regime 4, long range and high density impurities, $\xi \gg a$
and $\xi \gg R_i$.} In the case of low density impurities, the
scatterings due to different impurities are independent. But when
the density is sufficiently high, so that potential field induced by
different impurities overlap, and multi-scattering dominates. This
multi-scatterings have no obvious effect on the existence of the
dip. While in the high energy region, the linear relation
$\sigma(E_F)$ breaks down and the curve degenerates into a square
root like curve and $\sigma(V_g) \sim \sqrt[4]{V_g}$
correspondingly, see Fig. \ref{FxiD} (d).

Finally, we discuss the energy scaling of the impurity potential $W$
considered to be fixed thus far. In our calculations $W$ is much
larger than the level spacing of sub-bands. The opposite limit has
been investigated recently, and a perfectly conducting channel was
found \cite{Wa07}.

\emph{Conclusions.}---As a summary, we numerically investigate the
transport properties of graphene nanoribbons in the presence of
the impurities with different density and potential range. In the
Fermi energy region of focus, four typical types of behavior can
appear from the unconventional electronic structures in zigzag
graphene nanoribbons, which can be tested by future experiments.
The third regime for the low density and long range impurities can
be used to explain the nonlinearity of $\sigma(V_g)$ in a recent
experiment \cite{Miao07}.

We acknowledge useful discussions with Professors C. N. Lau, S. C.
Zhang, Q. Niu, C. W. J. Beenakker, J. R. Shi and Q. F. Sun. This
work was supported by NSF of China under grant 90406017, 60525417,
10610335, the NKBRSF of China under Grant 2005CB724508 and
2006CB921400. X. C. Xie is supported by US-DOE and US-NSF.
\bibliographystyle{plain}

\end{document}